\documentclass[twocolumn,aps,prb,amsmath,floatfix]{revtex4}
\usepackage{graphicx}
\begin{document}

\title{Single Mott Transition in Multi-Orbital Hubbard Model}
\author{A. Liebsch} 
\affiliation{Institut f\"ur Festk\"orperforschung, 
         Forschungszentrum J\"ulich, 
         52425 J\"ulich, Germany}
\begin{abstract}
The Mott transition in a multi-orbital Hubbard model involving 
subbands of different widths is studied within the dynamical mean
field theory. Using the iterated perturbation theory for the 
quantum impurity problem it is shown that at low temperatures 
inter-orbital Coulomb interactions give rise to a single first-order 
transition rather than a sequence of orbital selective transitions. 
Impurity calculations based on the Quantum Monte Carlo method confirm 
this qualitative behavior. Nevertheless, at finite temperatures, 
the degree of metallic or insulating behavior of the subbands differs 
greatly. Thus, on the metallic side of the transition, the narrow band 
can exhibit quasi-insulating features, whereas on the insulating side 
the wide band exhibits pronounced bad-metal behavior. This complexity
might partly explain contradictory results between several previous 
works.  
\end{abstract}
\maketitle

\section{Introduction}
The nature of the metal insulator transition in multi-band materials 
involving subbands of different widths is not yet fully understood. 
As a result of the geometric complexity of many transition metal 
oxides, the degeneracy of valence bands is frequently lifted, giving 
rise to coexisting narrow and wide partially filled bands. A key 
issue in these materials is therefore whether these subbands exhibit
separate Mott transitions or whether single-particle hybridization   
among subbands and inter-orbital Coulomb interactions ensure the
occurrence of a single transition involving all subbands simultaneously. 
An example is the layer perovskite Sr$_2$RuO$_4$ which consists 
of a wide, two-dimensional $d_{xy}$ band and narrow, nearly 
one-dimensional $d_{xz,yz}$ bands.\cite{oguchi} This system is 
believed to exhibit unconventional p-wave superconductivity. 
\cite{maeno} Iso-electronic replacement of Sr by Ca leads to 
a distortion of oxygen octahedra and an effective narrowing of 
the Ru-derived $t_{2g}$ bands, causing a metal insulator transition. 
\cite{nakatsuji} Another case is the layer compound Na$_x$CoO$_2$ 
which as a function of doping concentration exhibits a wide range 
of properties, including superconductivity (when hydrated), while 
the parent material CoO$_2$ is an insulator.\cite{louie}        

In previous work \cite{EPL} we investigated the Mott transition in 
Ca$_{2-x}$Sr$_x$RuO$_4$ by considering a 3-band model appropriate 
for Sr$_2$RuO$_4$ and varying the on-site Coulomb interaction
rather than modifying the band width. 
Extending earlier quasi-particle calculations for this system 
\cite{liebsch} within the dynamical mean field theory\cite{DMFT} 
(DMFT) we found that inter-orbital Coulomb interactions lead
to a significant redistribution of spectral weight between $t_{2g}$ 
orbitals. As a result, despite the completely different shapes and 
widths of the single-particle densities of states, for increasing 
$U$ the $d_{xy}$ and $d_{xz,yz}$ subbands exhibit similar correlation
features, suggesting the existence of a single Mott transition.     
Additional single-particle coupling between $t_{2g}$ states 
caused by octahedral distortions \cite{fang} undoubtedly will  
enhance the trend towards a single transition. This result was
in conflict with DMFT calculations by Anisimov et al.\cite{anisimov} 
who obtained a sequence of Mott transitions as the on-site Coulomb
interaction was increased beyond the single-particle widths of the
narrow and wide $t_{2g}$ subbands of Sr$_2$RuO$_4$, respectively.  

To avoid the uncertainties stemming from the maximum entropy 
reconstruction \cite{jarrell} of real-frequency spectra 
we recently considered a simple two-band model and focused on 
quantities directly available from the Quantum Monte Carlo (QMC) 
calculations carried out at imaginary times and frequencies.
\cite{prl} The subbands were assumed to have semi-circular 
density of states of widths $W_1=2$~eV, $W_2=4$~eV, to be 
each half-filled, and to interact only via local intra- and 
inter-orbital Coulomb energies $U$ and $U'=U-2J$, where $J$ 
is the Hund's rule exchange integral. The quantum impurity 
calculations were performed within the DMFT and multi-orbital 
QMC method for $T=125$~meV. For increasing $U$ and $U'$ (with 
fixed $J=0.2$~eV) the subband quasi-particle weights $Z_i$
were found to vanish at about the same critical $U_c$, within a
characteristic uncertainty caused by critical slowing down. 
The same behavior was found for the quantities $G_i(\beta/2)$  
($\beta=1/k_BT$) which define the spectral weight 
of the coherent subband peaks within a few $k_BT$ of $E_F$. 
Also, two-band calculations at $T=0$ within the iterated 
perturbation theory\cite{IPT} (IPT) supported the existence
of a common transition. These results confirmed our previous 
finding, namely, that inter-subband coupling due to local Coulomb 
interactions gives rise to a single Mott transition rather than 
a sequence of orbital-selective transitions. 

Recently Koga et al.\cite{koga} considered the same two-band 
Hubbard model involving semi-circular densities of states with 
$W_1=2$~eV, $W_2=4$~eV and on-site Coulomb interactions 
specified by $U=U'+2J$. Also assuming both subbands to be 
half-filled, they studied the effect due to larger Hund's rule
coupling by choosing $J=U/4$ and $U'=U/2$. To be able to reach 
lower temperatures, the DMFT calculations were performed 
within the exact diagonalization (ED) method \cite{ED} and
a linearized two-site version of the DMFT.\cite{potthoff}        
Surprisingly, both impurity treatments indicate the existence 
of orbital-selective metal insulator transitions, i.e., the 
narrow subband exhibits a smaller critical $U_c$ than the 
wide subband, implying a region of coexisting metallic and
insulating subbands in spite of the inter-orbital Coulomb 
interactions defined by $U'$ and $J$.

The conflicting results obtained within these various approaches
underline the subtle and complex nature of the Mott transition
in multi-orbital systems. Since the non-isotropic multi-band 
character among transition metal oxide materials is the rule 
rather than an exception there is evidently a need to clarify        
this fundamental issue. Further theoretical studies are required 
to test the reliability of the DMFT method and various quantum 
impurity treatments. Contradictory results for a rather simple 
model which nevertheless captures a key feature present in many 
real systems obviously shed doubts on predictions obtained for 
more complex materials. 

In the present work we reconsider the same two-band Hubbard model 
as in Refs.~11 and 13. Since the QMC calculations 
are difficult to perform at low temperatures, we consider first the 
simpler iterated perturbation theory at finite $T$. Although this
scheme is not reliable on a quantitative level, it can serve as a
useful guide for qualitative purposes. At sufficiently low 
temperatures, the IPT reveals a single first-order transition 
for both subbands, confirming our previous results. 
Because of the numerical simplicity of this model we are able to 
obtain the entire $T-U$ phase diagram. We then study the same model 
within the two-band QMC method. Although convergence is considerably 
more difficult at low temperatures, these results also suggest a 
common Mott transition. On the other hand, the interpretation of 
this multi-band phase transition is complicated since in both 
phases the subbands exhibit quite different correlation features 
due to their different single-particle properties. Thus, in the 
metallic phase, the narrow band is much more correlated than the 
wide band and shows a mixture of both metallic and insulating features. 
Similarly, in the insulating phase the narrow band exhibits a gap 
while the wide band reveals pronounced bad-metal behavior. Thus, 
the transition is partially incomplete for individual subbands.
This complexity of the quasi-particle spectra, and the difficulty of 
clearly identifying metallic and insulating properties at finite
temperatures, presumably is the origin of some of the contradictions 
between the previous works.     
 
The outline of this paper is as follows. In the next section we
discuss the DMFT results derived within the IPT as a function of
temperature. Section III contains the results obtained for the
QMC method. Section IV provides a summary. 

\section{IPT -- DMFT}

Let us consider the paramagnetic metal insulator transition in a 
two-band Hubbard model consisting of subbands of width $W_1=2$~eV and 
$W_2=4$~eV. The densities of states are assumed to be semi-circular:
\,$\rho_i(\omega)=4/(\pi W_i)(1 - 4\omega^2/W_i^2)^{1/2}$, 
corresponding to non-hybridizing Bethe lattices. Both bands are 
taken to be half-filled. The interacting Green's functions at 
imaginary frequencies are defined as
\begin{equation}
  G_i(i\omega_n)= \int\!d\omega\ \frac{\rho_i(\omega)}
                  {i\omega_n + \mu - \omega - \Sigma_i(i\omega_n)}\ , 
\end{equation} 
where $\omega_{n\ge0}=(2n+1)\pi k_BT$ are Matsubara frequencies and
$\mu$ is the chemical potential.
$\Sigma_i(i\omega_n)$ are the subband self-energies which must be
determined self-consistently. Removal of the self-energy from the 
central site yields the impurity Green's functions 
\begin{equation}
   g_i(i\omega_n)=[G_i^{-1}(i\omega_n) + \Sigma_i(i\omega_n)]^{-1}\ . 
\end{equation}
The corresponding Fourier transforms at imaginary times are 
denoted as $g_i(\tau)$.
Since both bands are assumed to be symmetric and half-filled, 
these functions satisfy the conditions 
\,$g_i(0)=g_i(\beta)=-0.5$\, and \,$g_i(\tau)=g_i(\beta-\tau)$. 
For purely on-site Coulomb interactions specified by intra- and
inter-orbital matrix elements, the self-energy components in 
second-order perturbation theory are:\cite{liebsch}
\begin{equation}
  \Sigma_i(i\omega_n) =  \int_0^\beta\!d\tau\, e^{i\omega_n\tau}\,
     [\sigma_1 g_i^3(\tau) + \sigma_2 g_i(\tau) g_j^2(\tau)]\ ,
\end{equation} 
where $j=1(2)$ for $i=2(1)$ and the coefficients $\sigma_i$ are 
defined as \,$\sigma_1 = U^2$ and \,$\sigma_2 = J^2 + 2(U'^2 +J^2 -U'J)$.
Since we take both subbands to be half-filled we do not consider 
here corrections to the bare second-order self-energy which should 
be introduced in the absence of particle-hole symmetry.
\cite{kajueter} 
At half-filling, the first-order Hartree-Fock terms are identical 
for both bands, $\Sigma_i^{\rm HF}=1.5U-2.5J$, and can be absorbed 
by the chemical potential. Throughout this paper we assume $J=U/4$ 
and $U'=U/2$, unless noted otherwise. 

The self-energies $\Sigma_i(i\omega_n)$ are now inserted in Eq.(1) 
and the procedure is iterated until self-consistency is achieved. 
The starting self-energy is taken to be zero. In particular at low 
temperatures, the direct and inverse Fourier transforms must be done 
with sufficient accuracy.\cite{victor} Also, at each iteration the 
symmetry condition \,$g_i(\tau)=g_i(\beta-\tau)$\, is enforced 
to remove spurious effects due to numerical inaccuracies. This 
ensures that the $\Sigma_i(i\omega_n)$ remain purely imaginary 
as they should be for symmetric half-filled bands. Typically
$L=256$ time slices and $2^{12}$ Matsubara frequencies were used. 
Also, the real-$\omega$ integration in Eq.\,(1) must be performed 
on a rather fine mesh in order to capture the narrow coherent peak
close to the metal insulator transition. This can be achieved by
linearly interpolating $\rho_i(\omega)$ and performing the $\omega$
integral analytically between mesh points. Once a converged solution
is found, the Green's functions $G_i(i\omega_n)$ are used to generate 
the Fourier transforms $G_i(\tau)$ from which the quasi-particle 
spectra at real frequencies are derived via the maximum entropy 
method.\cite{jarrell}   

\begin{figure}[t!]
  \begin{center}
  \includegraphics[width=4.9cm,height=7cm,angle=-90]{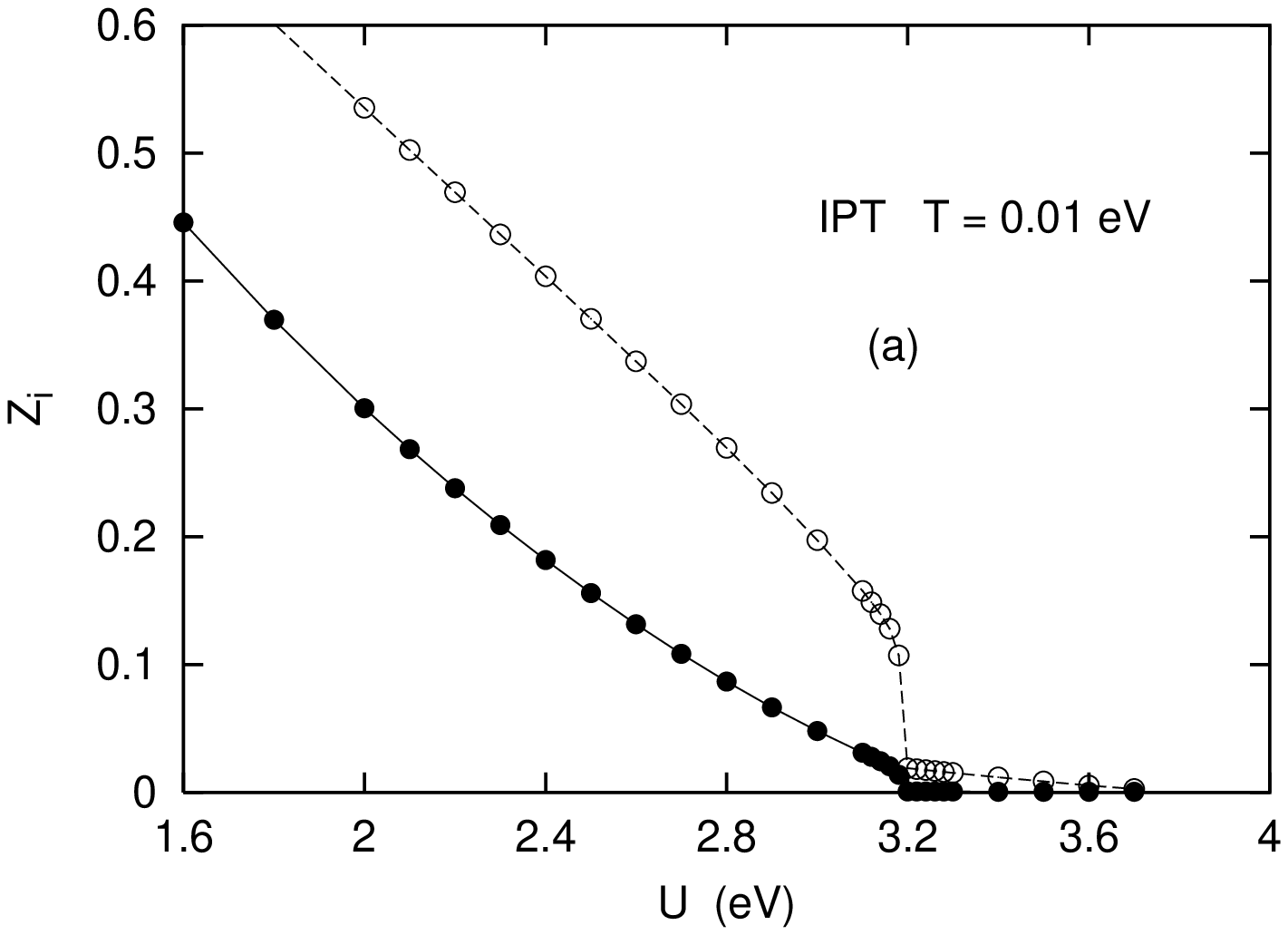}
  \includegraphics[width=4.9cm,height=7cm,angle=-90]{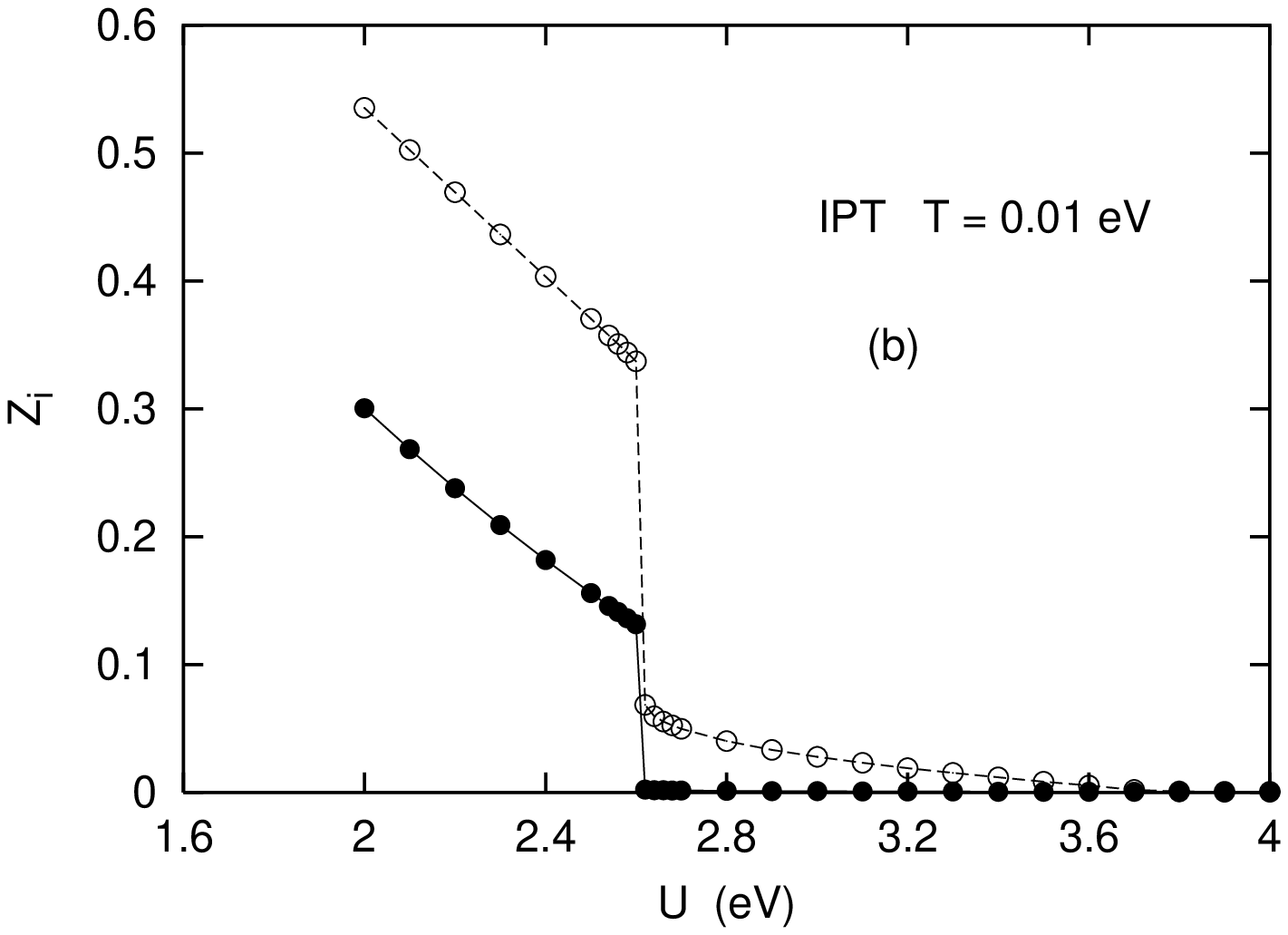}
  \end{center}
  \vskip-3mm
\caption{
Quasi-particle weights $Z_i$ of two-band Hubbard model as a function of 
on-site Coulomb energy $U$ for $T=0.01$~eV, calculated within IPT-DMFT.
Solid (open) dots: $Z_1$ ($Z_2$). (a) increasing $U$, (b) decreasing $U$.  
}\end{figure}

\begin{figure}[t!]
  \begin{center}
  \includegraphics[width=4.9cm,height=7cm,angle=-90]{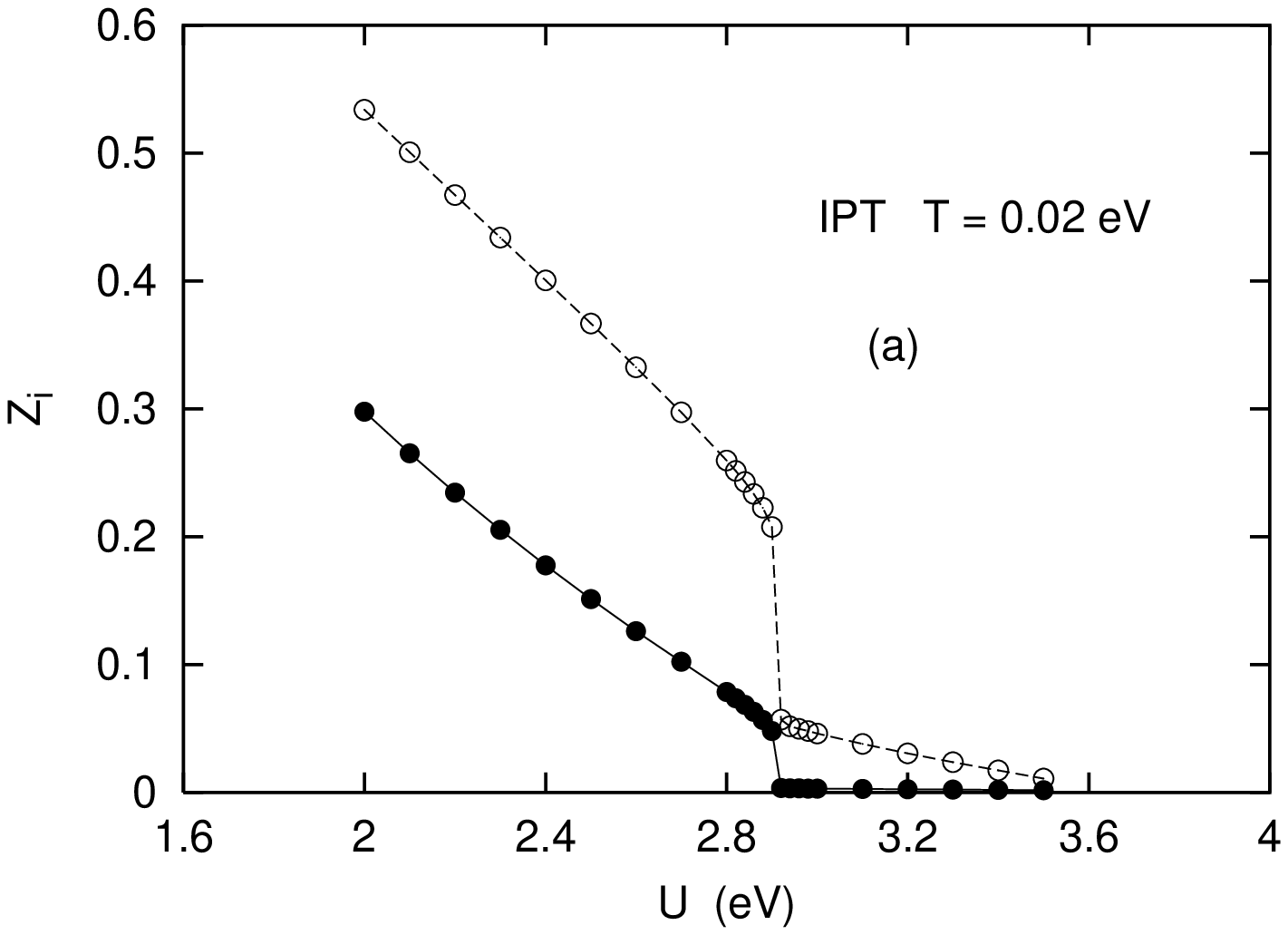}
  \includegraphics[width=4.9cm,height=7cm,angle=-90]{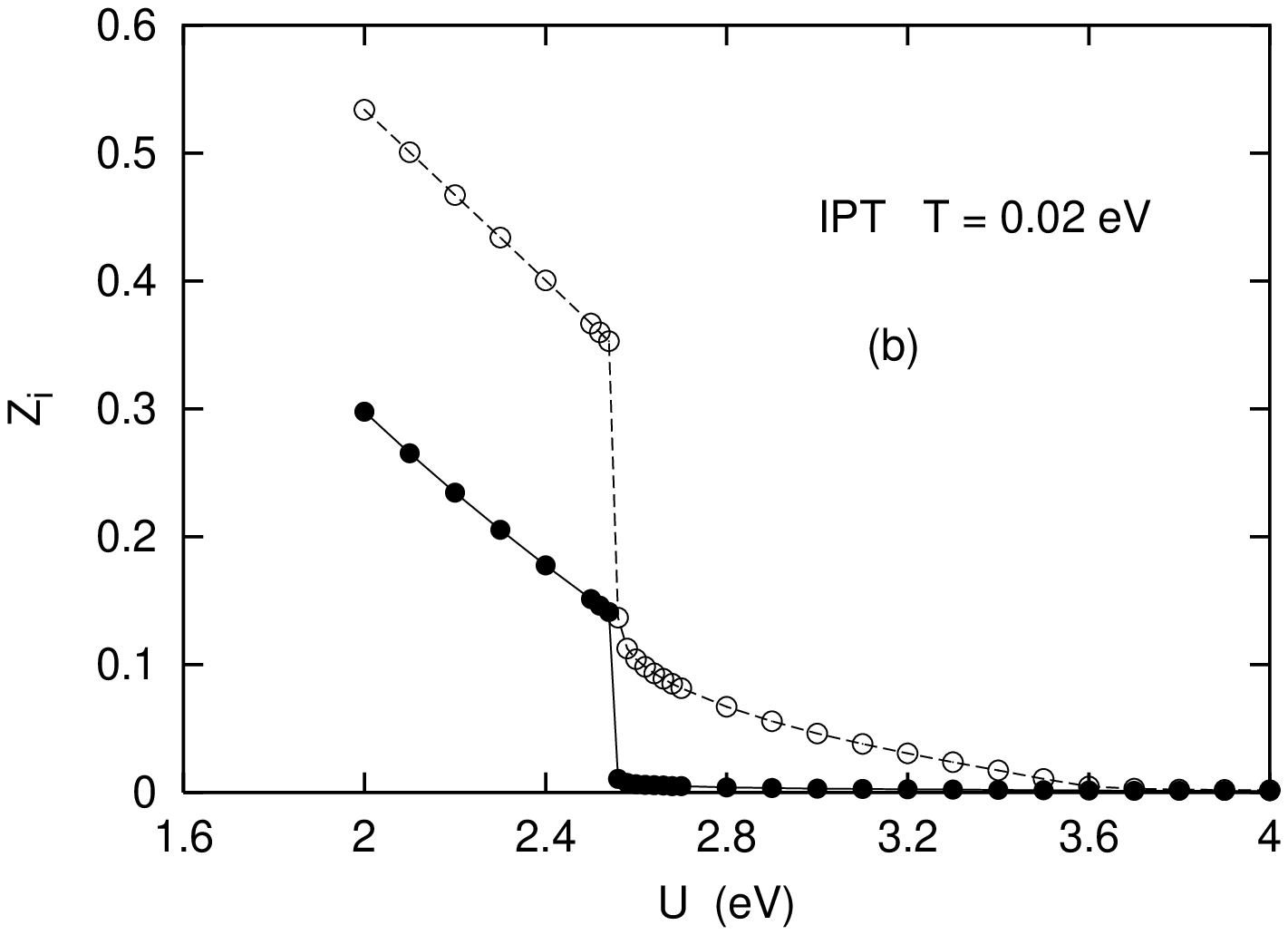}
  \end{center}
   \vskip-3mm
\caption{
Same as Fig.~1 except for $T=0.02$~eV.
}\end{figure}

\begin{figure}[t!]
  \begin{center}
  \includegraphics[width=4.9cm,height=7cm,angle=-90]{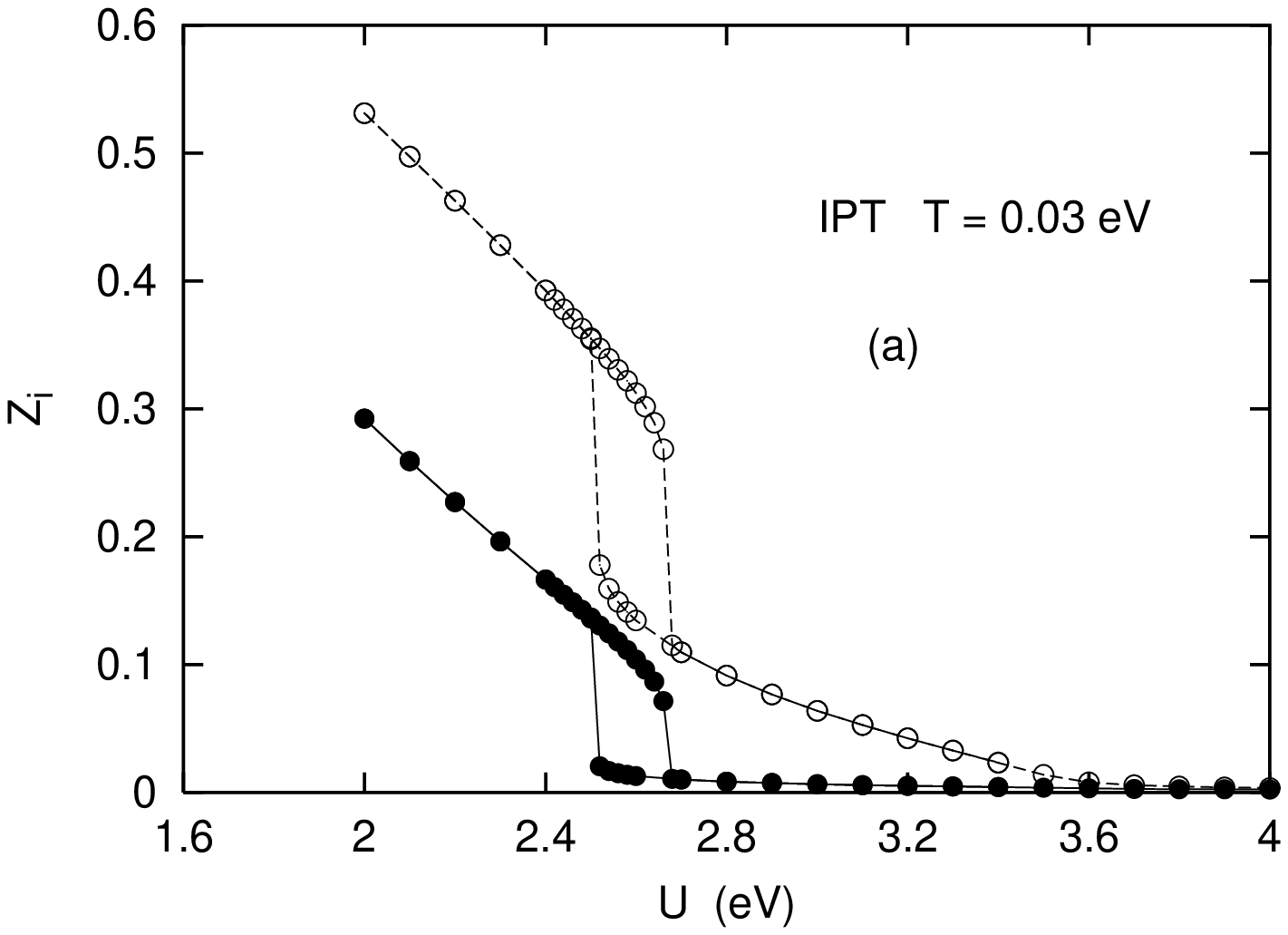}
  \includegraphics[width=4.9cm,height=7cm,angle=-90]{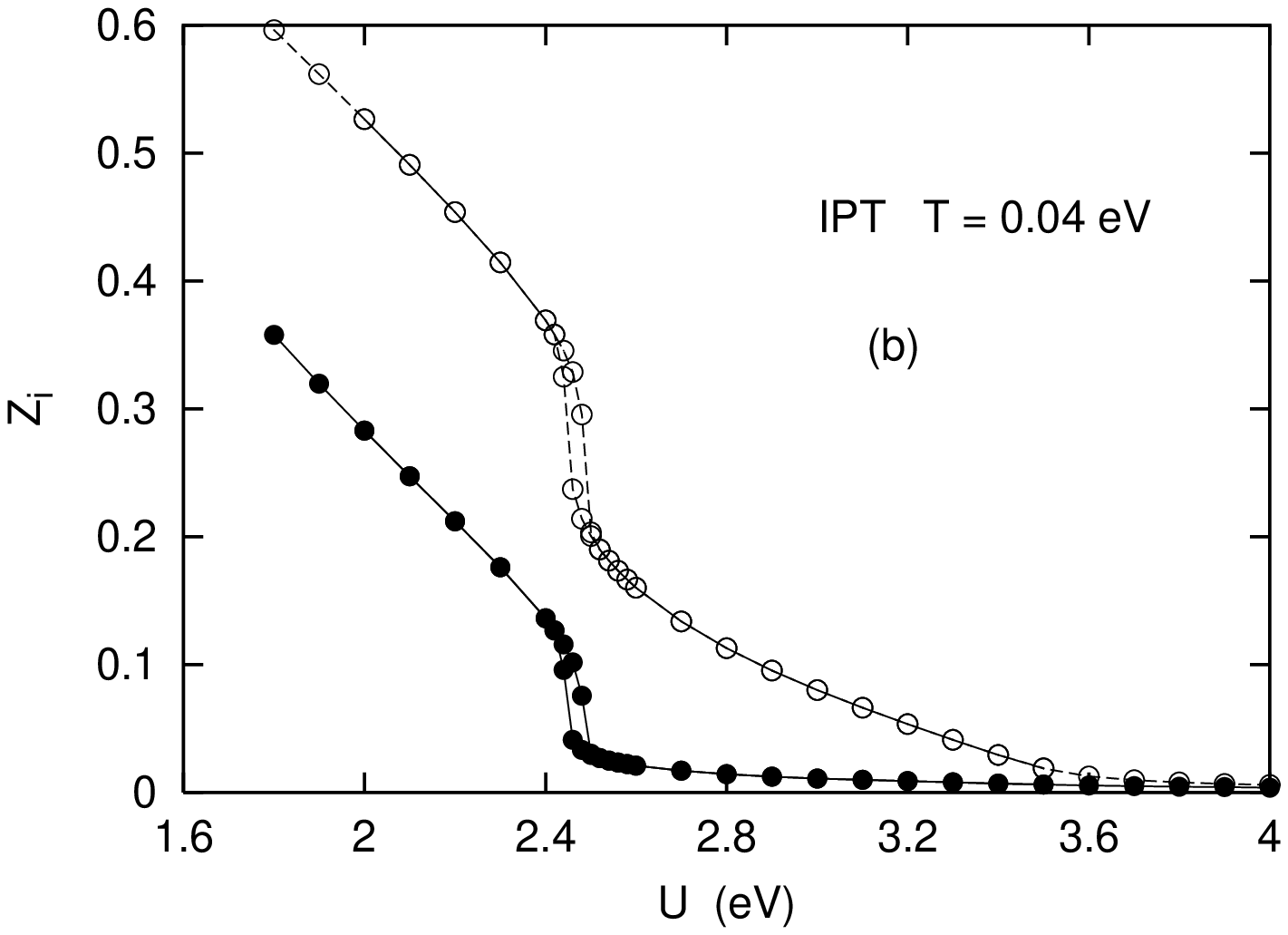}
  \end{center}
   \vskip-3mm
\caption{
Same as Fig.~1 except for (a) $T=0.03$~eV and (b) $T=0.04$~eV.
Each Figure shows the hysteresis obtained for increasing $U$ 
(upper points) and decreasing $U$ (lower points).  
}\end{figure}

\begin{figure}[t!]
  \begin{center}
  \includegraphics[width=4.9cm,height=7cm,angle=-90]{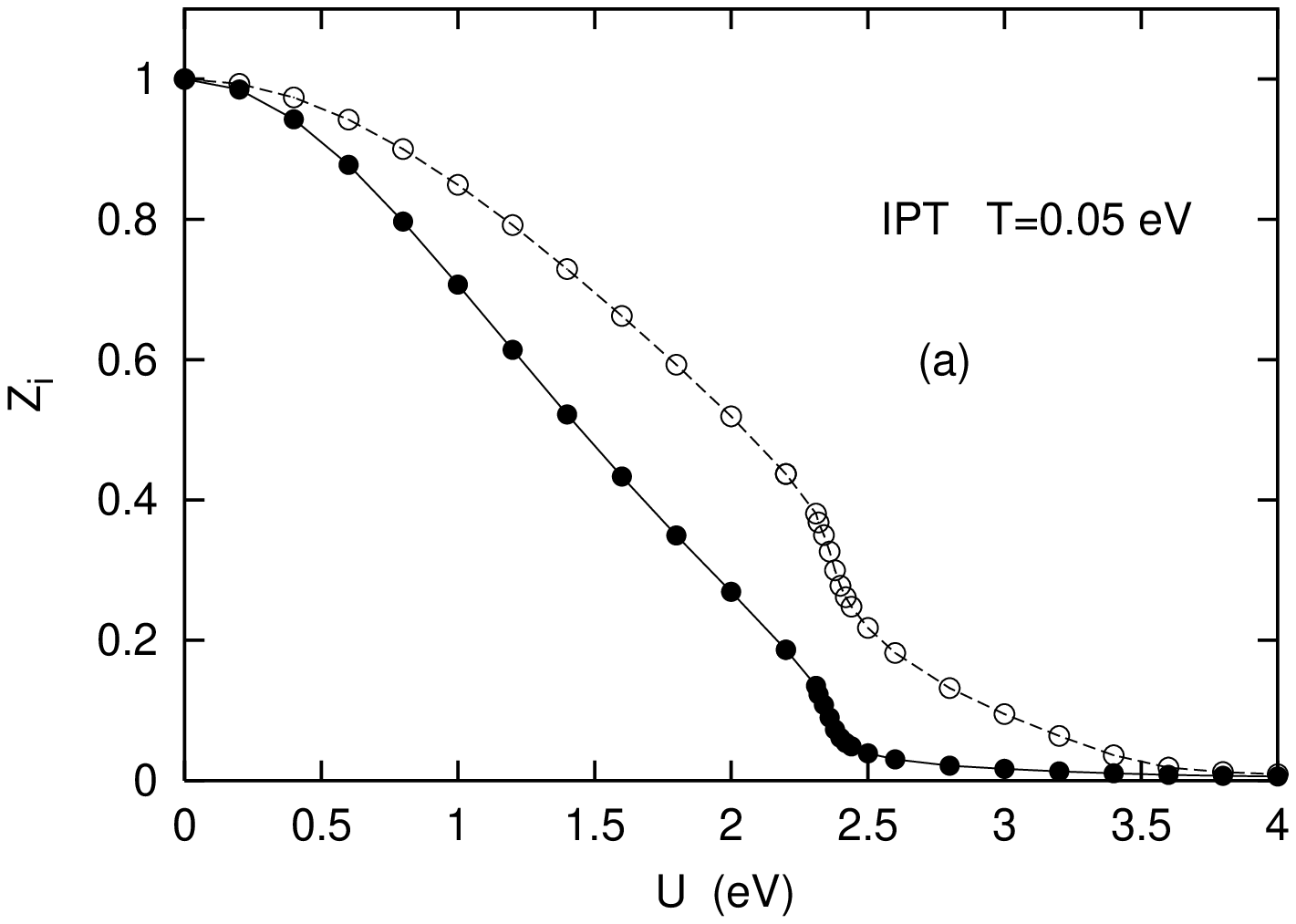}
  \includegraphics[width=4.9cm,height=7cm,angle=-90]{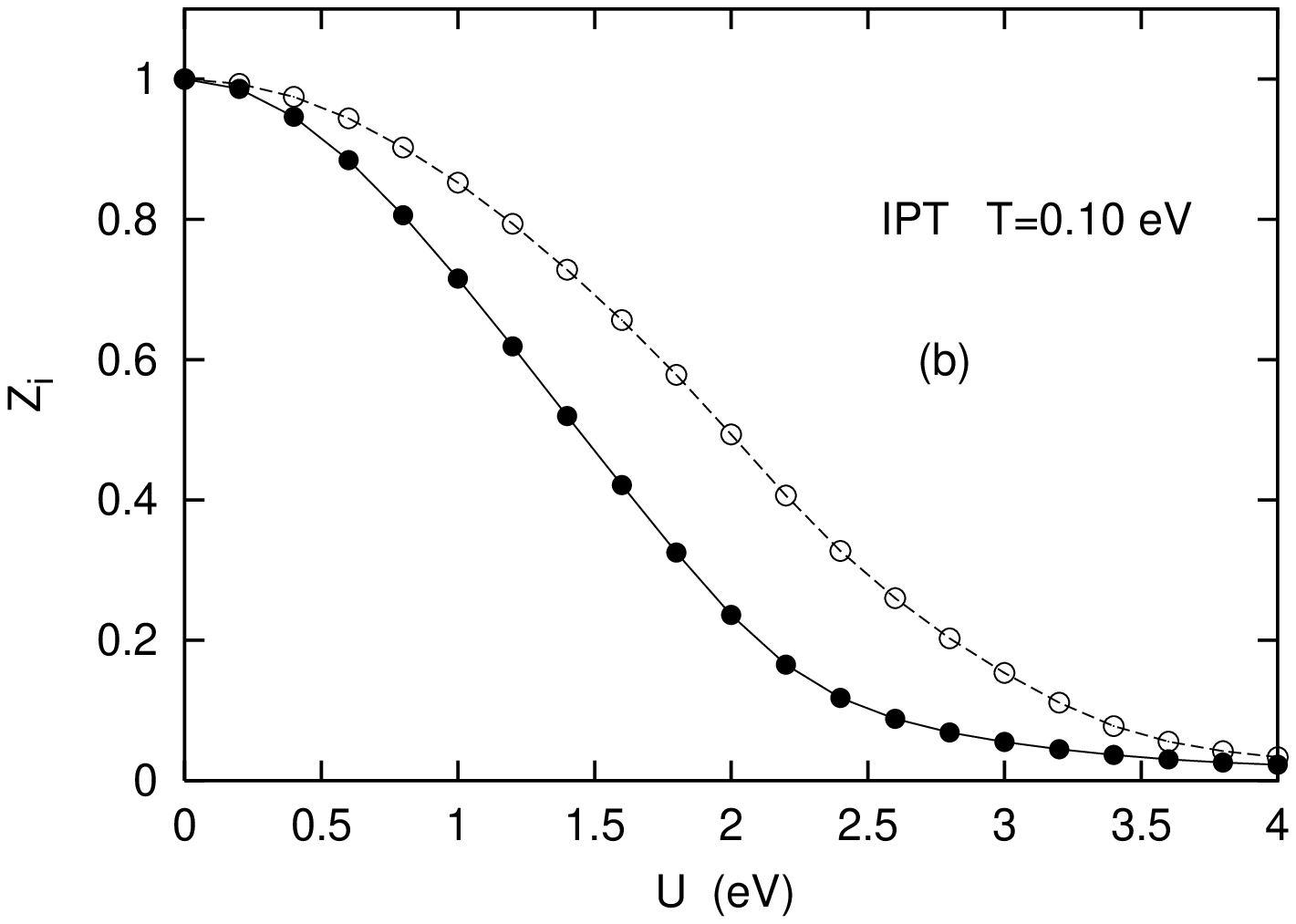}
  \end{center}
   \vskip-3mm
\caption{
Same as Fig.~1 except for (a) $T=0.05$~eV and (b) $T=0.1$~eV.
Hysteresis effects are negligible at these temperatures.
}\end{figure}

Figures 1 to 4 show the variation of the subband quasi-particle 
weights \,$Z_i\approx 1/[1-{\rm Im}\,\Sigma_i(i\omega_0)/\omega_0]$\, 
as functions of $U$ for various temperatures. Up to $T=0.04$~eV 
the results show the typical hysteresis behavior obtained also for 
the single-band Hubbard model.\cite{DMFT,victor,bulla} If $U$ is 
gradually increased from the metallic side the transition occurs 
at a slightly larger critical value than if one begins in the 
insulating phase and gradually decreases $U$. The hysteresis is most
pronounced at low temperatures and becomes weaker as $T$ increases. 
For $T\geq0.05$~eV it is negligibly small and the transition begins 
to resemble the typical cross-over behavior.

The important point of these results is that the quasi-particle 
weights of the narrow and wide subbands exhibit first-order
transitions at precisely the same critical $U_c(T)$. This picture
supports our previous results suggesting a single Mott transition 
in a non-isotropic multi-orbital environment, in contrast to the
orbital selective transitions found in Refs.~9 and 11.
In line with the notation used in the single-band case, we denote 
the common critical value for decreasing $U$ as $U_{c1}(T)$ and 
for increasing $U$ as $U_{c2}(T)$. (These values should not be 
confused with the critical $U$ derived in Ref.~11 for the 
isotropic two-band models.) For $T\leq0.04$~eV the transitions 
are perfectly discontinuous both at $U_{c1}(T)$ and $U_{c2}(T)$. 
For $T\geq 0.05$~eV finite temperature broadening progressively
dominates the transition region.

\begin{figure}[t!]
  \begin{center}
  \includegraphics[width=5.0cm,height=8cm,angle=-90]{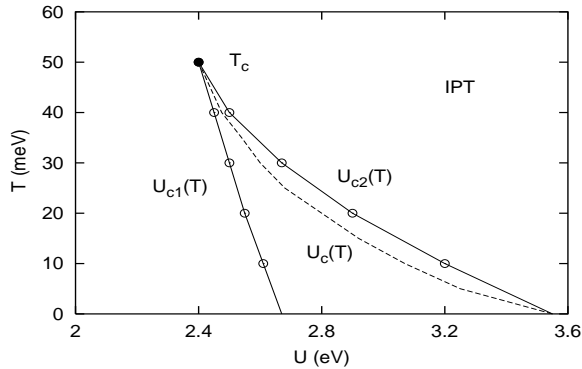}
  \end{center}
\caption{
Phase diagram for two-band Hubbard model, calculated within IPT-DMFT.
Open dots: critical Coulomb energies $U_{c1}(T)$ and $U_{c2}(T)$ 
derived from Figs.~1 to 4. The lines are guides to the eye. Their 
extrapolation yields the common second-order critical point marked 
by a solid dot. The actual first-order line $U_{c}(T)$ at low 
temperatures lies between $U_{c1}(T)$ and $U_{c2}(T)$ and is
sketched by the dashed line.    
}\end{figure}

Figure 5 shows the phase diagram obtained from the functions $Z_i(T,U)$. 
The overall shape of the phase boundaries agrees well with those 
obtained for the single-band Hubbard model.\cite{DMFT,bulla}
According to the single-band phase diagram, the actual first-order 
metal insulator transition takes place along an intermediate line 
$U_c(T)$ such that \,$U_{c1}(T)<U_{c}(T)<U_{c2}(T)$
(schematically indicated by the dashed curve).
Extrapolation of the phase boundaries suggests a common transition 
temperature of about $T_c\approx0.05$~eV. This value agrees well
with the average between the critical temperatures derived 
for isotropic narrow and wide two-band systems, 
$T_{c11}\approx0.033$~eV and 
$T_{c22}\approx0.067$~eV, respectively. 
[In the single band case the IPT-DMFT yields $T_{c}/W\approx0.02$.
\cite{DMFT} According to Eq.(3), for doubly degenerate bands and 
$U=2U'=4J$, this value should be reduced by a factor 
\,$(\sigma_1+\sigma_2)^{1/2}/U\approx1.2$.]      

Although inter-orbital Coulomb interactions lead to a common metal 
insulator transition in the narrow and wide subbands, it is clear 
from Figs.~1 to 4 that the subband quasi-particle weights differ 
greatly both in the metallic and insulating phases. In this sense,
the subband transitions are partly incomplete: On the metallic
side, $Z_1$ is consistently smaller than $Z_2$ because of the more 
pronounced correlations within the narrow band. As as result, as 
will be discussed below, this band can exhibit quasi-insulating 
behavior. Conversely, on the insulating side of the transition, 
$Z_1$ drops almost to zero while $Z_2$ drops to a finite 
intermediate value and gradually decreases towards larger $U$. 
This `bad-metal' behavior of the wide band above the transition 
is weak at low temperatures but becomes more important towards 
$T_c$. There is no indication of a second first-order transition 
in the wide subband at larger $U$. 

\begin{figure}[t!]
  \begin{center}
  \includegraphics[width=5.0cm,height=8cm,angle=-90]{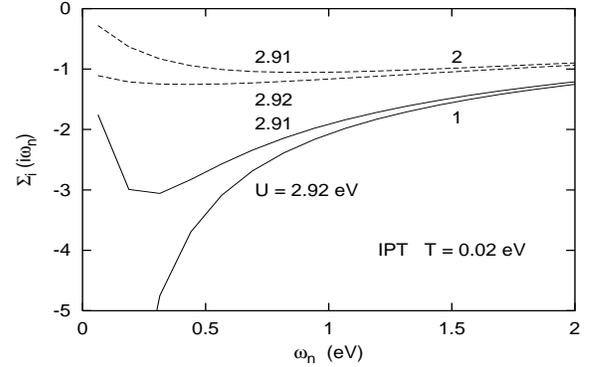}
  \end{center}
\caption{
Self-energy $\Sigma_i(i\omega_n)$ for two-band Hubbard model 
at $T=0.02$~eV near $U_{c2}(T)\approx2.91$~eV at small Matsubara 
frequencies. Solid (dashed) curves: narrow (wide) subband.
}\end{figure}

To analyze the subband quasi-particle properties close to the transition
in more detail we show in Fig.~6 the self-energies $\Sigma_i(i\omega_n)$ 
for $T=0.02$~eV near $U_{c2}(T)$. Slightly below the transition at 
$U=2.91$~eV the wide band is clearly metallic with $\Sigma_2(i\omega_n)
\sim\omega_n$ at small Matsubara frequencies. The narrow band exhibits 
significant deviations from this linear behavior. Only at the two lowest 
frequencies $\Sigma_1(i\omega_n)$ is roughly linear in $\omega_n$.
At larger frequencies it becomes inversely proportional to $\omega_n$. 
Thus, this band is in the intermediate region between metallic and 
insulating behavior, exhibiting a narrow coherent peak near $E_F$ nearly
separated from pronounced Hubbard bands (see below). At $U=2.92$~eV, 
i.e., just above the transition, the narrow band has become fully 
insulating with $\Sigma_1(i\omega_n)\sim1/\omega_n$. In contrast,
$\Sigma_2(i\omega_n)$ approaches a constant in the limit of small 
$\omega_n$. Thus, the wide band still is in the intermediate range 
between metallic and insulating behavior. These results demonstrate
that in the vicinity of the Mott transition the two subbands exhibit
a complicated superposition of different cross-over behaviors as 
they switch between the metallic and insulating phases. 

\begin{figure}[t!]
  \begin{center}
  \includegraphics[width=4.9cm,height=8cm,angle=-90]{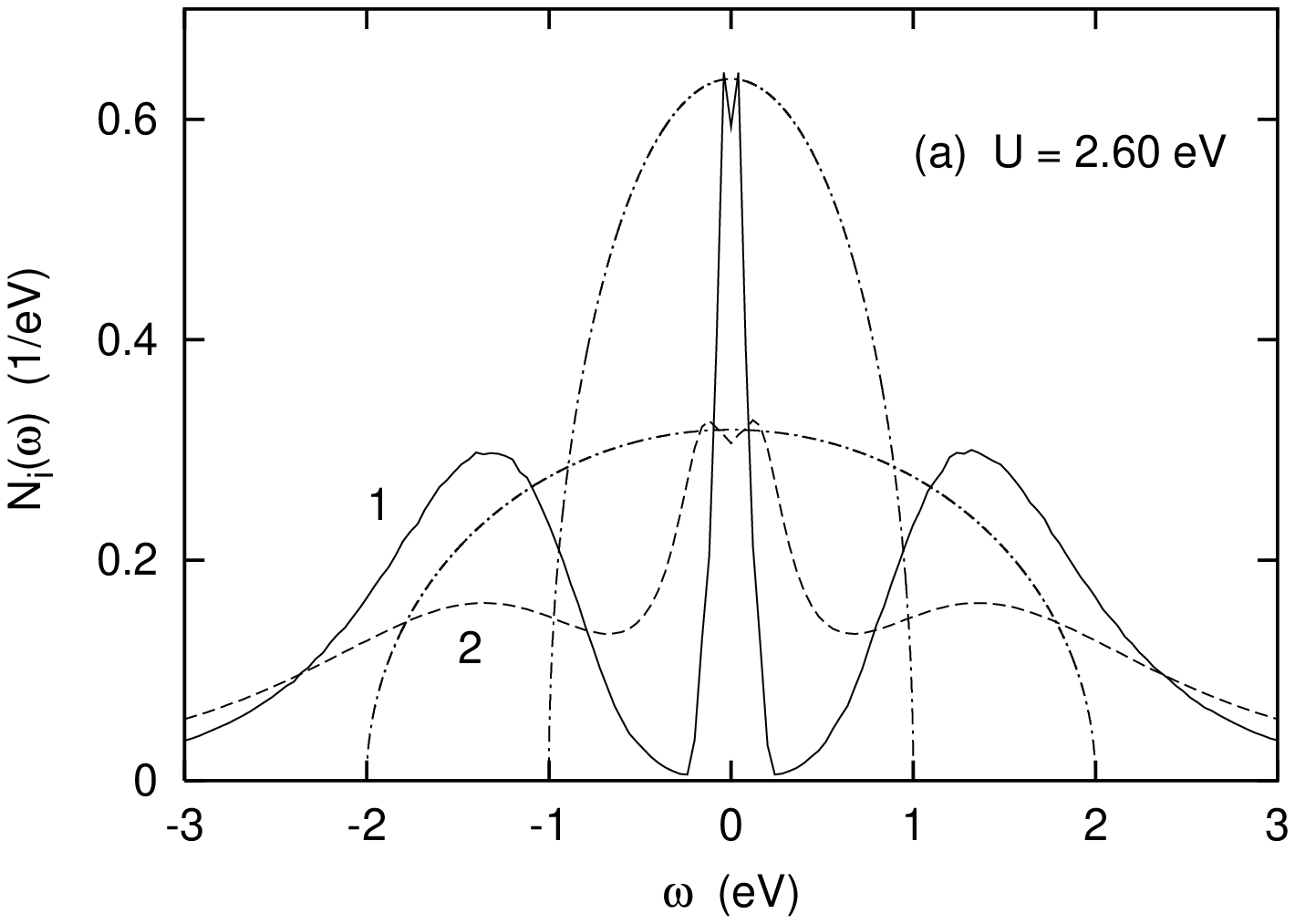}
  \includegraphics[width=4.9cm,height=8cm,angle=-90]{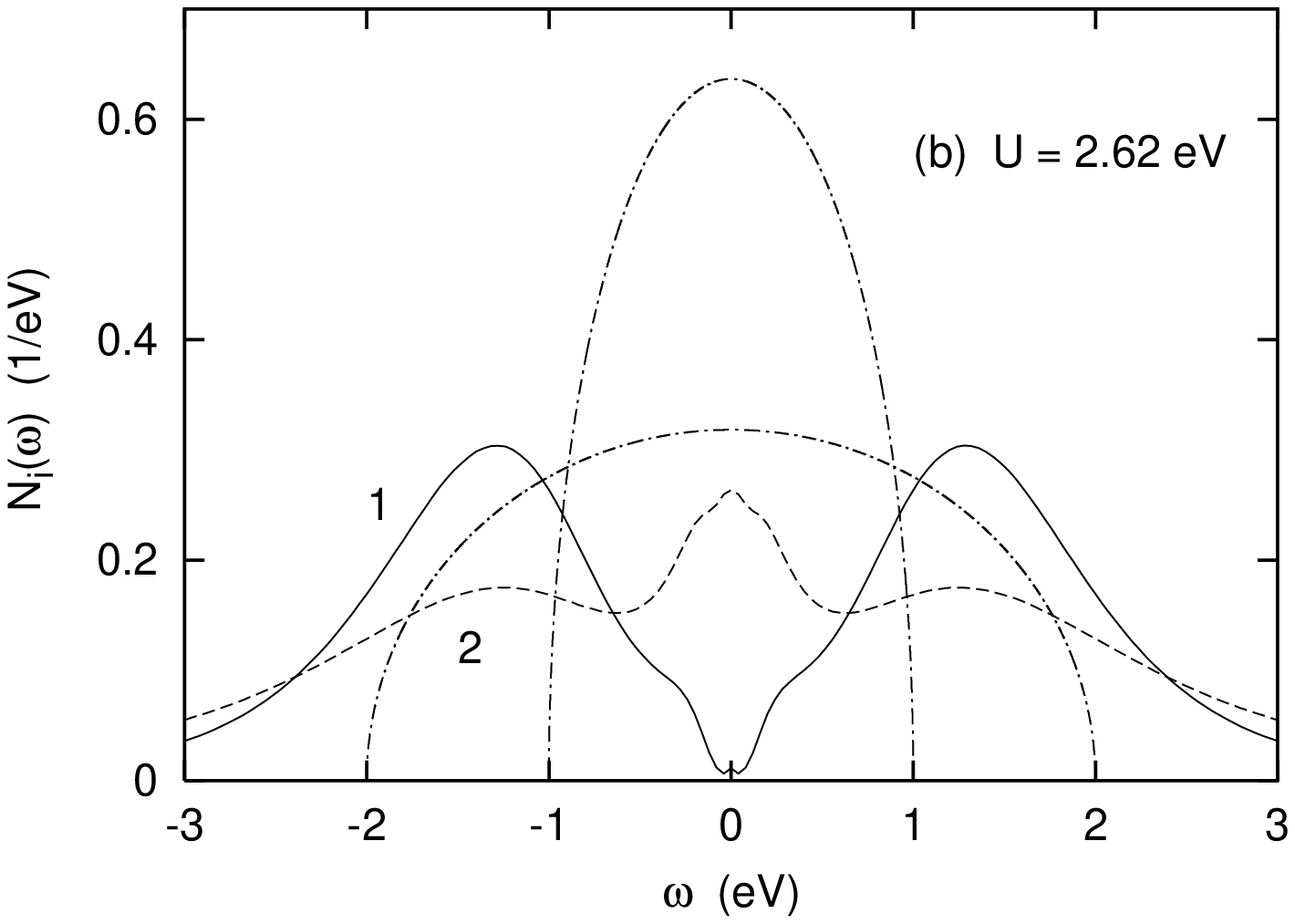}
  \end{center}
\caption{
Quasi-particle spectra $N_i(\omega)$ for two-band Hubbard model 
at $T=0.01$~eV near $U_{c1}(T)$.
(a) below the transition: $U=2.60$~eV; (b) above the transition: 
$U=2.62$~eV. Solid (dashed) curves: narrow (wide) subband; 
dash-dotted curves: bare densities of states. 
}\end{figure}

The effect of these subband dependent correlations on the 
quasi-particle spectra is illustrated in Fig.~7 for $T=0.01$~eV 
just below and above $U_{c1}(T)$. On the metallic side, the wide
band is perfectly metallic. The narrow band exhibits a very narrow 
coherent peak near $E_F$ and nearly vanishing weight between this 
peak and the large Hubbard peaks.\cite{dip} Thus, the narrow band 
shows features indicative of both metallic and insulating 
behavior. Similarly, above the transition the coherent peak in 
$N_1(\omega)$ has vanished and a narrow excitation gap is visible.
$N_2(\omega)$, on the other hand, still has appreciable spectral
weight near the Fermi energy. According to the phase diagram, at 
sufficiently low temperatures the true first-order line $U_c(T)$ 
approaches the boundary $U_{c2}(T)$. In fact, within a consistent 
dynamical two-site approximation \cite{potthoff2} $U_c(T)$ is only 
slightly smaller than $U_{c2}(T)$. Fig.~1 shows that near this 
line the `bad-metal' character of the wide subband above the 
transition is much weaker than close to $U_{c1}(T)$. 

The above results suggest that, although there is a single metal 
insulator transition for both subbands, slightly below and above 
this transition the quasi-particle spectra are rather complicated, 
partially exhibiting `coexisting' metallic and insulating behavior. 
In fact, in the range $0.01$~eV~$\leq T \leq T_c$, the transition 
in the wide band might be more appropriately described as 
metal/bad-metal transition. However, this behavior is associated 
with the fact that at finite $T$ there is no clear distinction 
between metal and insulator. Extrapolation of the results shown 
in Figs.~1 to 4 to even lower temperatures suggests that the 
`bad-metal' behavior in the wide band above $U_{c2}(T)$ eventually 
disappears and that the gaps in both bands open simultaneously 
at the same critical $U_c(T)$.   

 \begin{figure}[t!]
  \begin{center}
  \includegraphics[width=5.0cm,height=8cm,angle=-90]{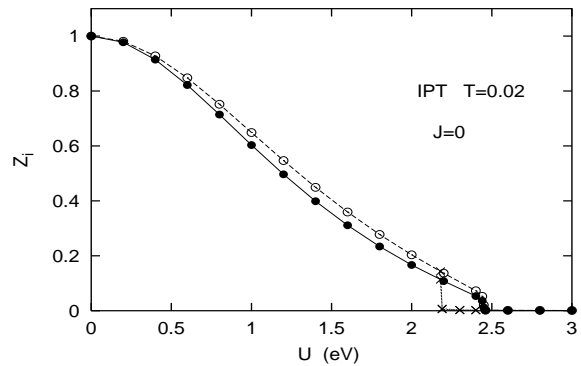}
  \end{center}
\caption{
Quasi-particle weights $Z_i$ of two-band Hubbard model as a function 
of on-site Coulomb energy $U$ for $T=0.02$~eV and $J=0$. Solid (open) 
dots: $Z_1$ ($Z_2$) for increasing $U$. Symbols (x) in the range 
$2.2\ldots2.5$~eV: $Z_1, Z_2\approx 0$ for decreasing $U$.
}\end{figure}

Before closing this section we briefly address the effect of the 
Hund's rule exchange coupling on the correlations within the 
two-band Hubbard model. The relatively large value assumed so far,
$J=U/4$, gives rise to considerable anisotropy and is partly the
origin of the qualitative differences between the subbands, in
particular near the transition. The reason for these differences 
can be understood in terms of the self-energy expression given in
Eq.~(3). For $J=U/4$: $\sigma_2/\sigma_1\approx0.5$ while
for $J=0$: $\sigma_2/\sigma_1=2$. Near the Fermi level
the single-particle density of states scales with the band width,
i.e., $g_1(\beta/2)=2g_2(\beta/2)$ before the Coulomb interaction
is switched on. Thus for $J=U/4$, $\tau=\beta/2$ one finds:
$\Sigma_1/\Sigma_2=
 (\sigma_1 g_1^3 + \sigma_2 g_1 g_2^2)/
 (\sigma_1 g_2^3 + \sigma_2 g_2 g_1^2)
 = (8 + 0.5\times 2)/(1 + 0.5\times 4)= 3 $
whereas for $J=0$: 
$\Sigma_1/\Sigma_2\approx(8 + 2\times 2)/(1 + 2\times 4)=4/3$. 
The subband contributions to the self-energy therefore differ 
significantly for $J=U/4$, but they are quite similar for $J=0$, 
giving rise to much more isotropic behavior.

Fig.~8 shows the quasi-particle 
weights $Z_i$ as functions of $U$ for $J=0$ at $T=0.02$~eV. 
The comparison with Fig.~2 demonstrates that as a result of the 
greater isotropy of the system the correlation induced reduction 
of quasi-particle weight is now quite similar for both subbands. 
In particular, slightly below and above the transition 
both bands show similar metallic and insulating behavior,
respectively, in contrast to the much more complex superposition 
of quasi-metallic and -insulating behavior obtained for $J=U/4$. 
As before, both bands undergo a first-order metal insulator
transition at the same critical $U$. Since there is neither 
single-particle hybridization nor Hund's rule coupling between
subbands, the common transition is entirely caused by the 
inter-orbital Coulomb interaction $U'=U$. Without this interaction,
we would have \,$Z_1(U) \approx Z_2(2U)$ (slight differences
arise as long as the temperature is held fixed). Thus, the 
critical Coulomb energies of the subbands would differ by the
same factor as the band widths $W_i$. Fig.~8 shows that $U'$ 
enforces a single transition at an intermediate $U_c$. 
As in Fig.~2, the $Z_i$ exhibit similar hysteresis behavior: 
for decreasing $U$ the critical Coulomb energy is 
$U_{c1}(T)\approx2.19$~eV while for increasing $U$  we find 
$U_{c2}(T)\approx2.45$~eV.  

\section{QMC -- DMFT}

In our previous work\cite{prl} we considered the same two-band
Hubbard model as above with \,$W_2=2W_1=4$~eV. DMFT calculations 
based on the multi-band QMC method were performed for a rather 
high temperature, $T=0.125$~eV, which corresponds to the cross-over 
region above the critical temperature $T_c$. For $J=0.2$~eV, the 
quasi-particle weights $Z_i$ were found to differ significantly 
at intermediate Coulomb energies due to the different subband 
kinetic energy terms. Nevertheless, with increasing $U$ the $Z_i$ 
merge again and, within some uncertainty associated with critical 
slowing down, become very small at about the same $U_c(T)$. 

\begin{figure}[t!]
  \begin{center}
  \includegraphics[width=5.0cm,height=8cm,angle=-90]{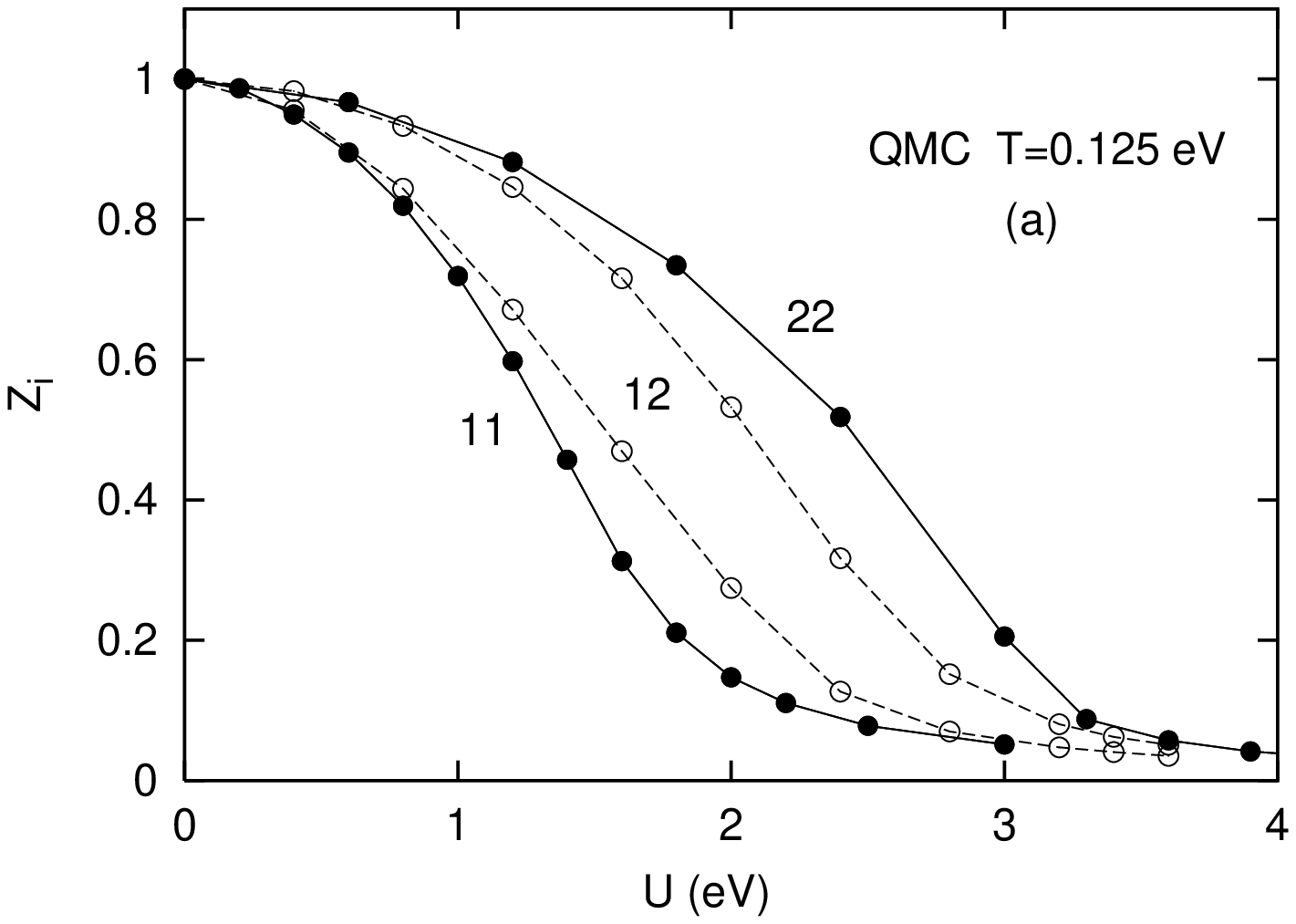}
  \includegraphics[width=5.0cm,height=8cm,angle=-90]{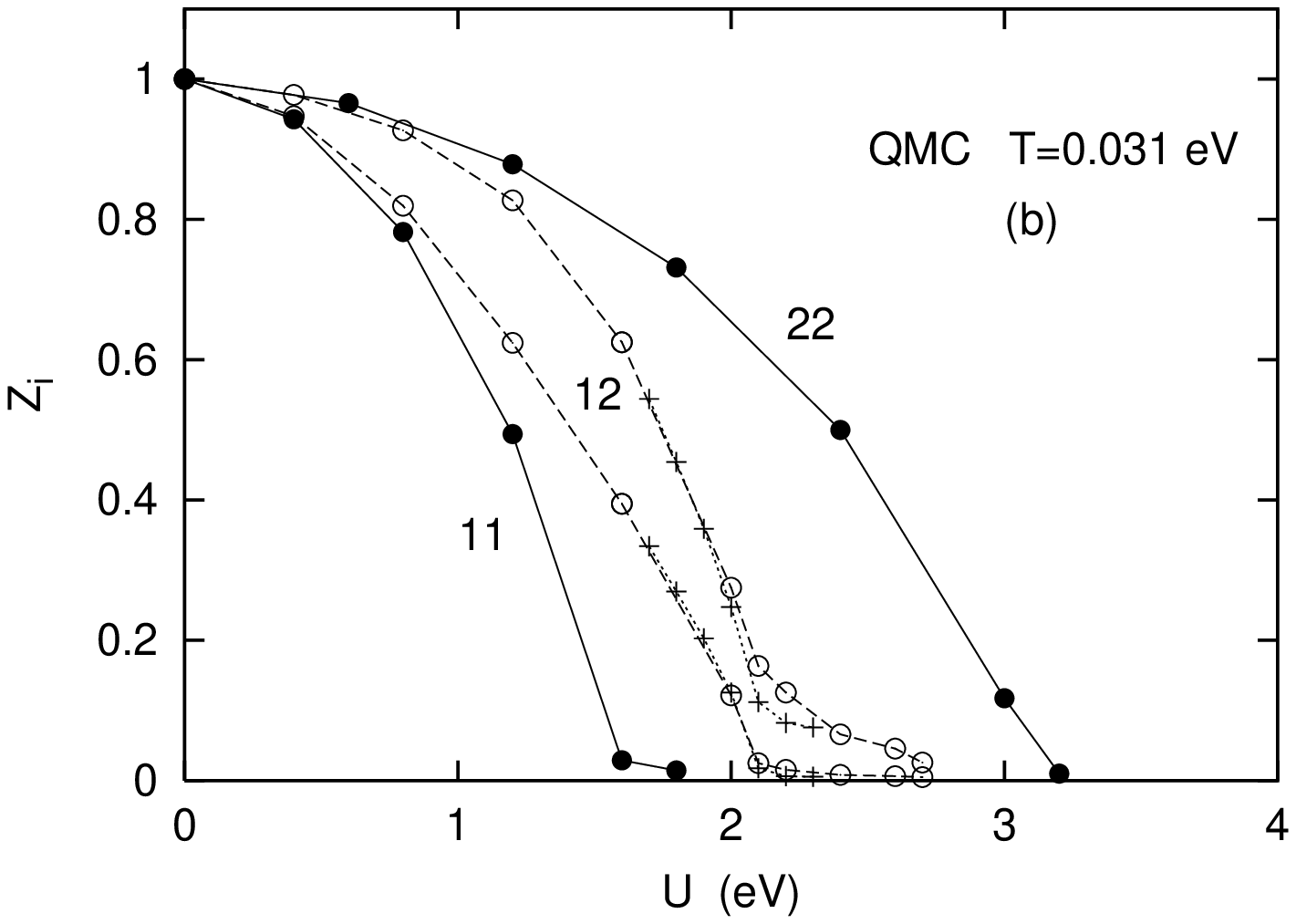}
  \includegraphics[width=5.0cm,height=8cm,angle=-90]{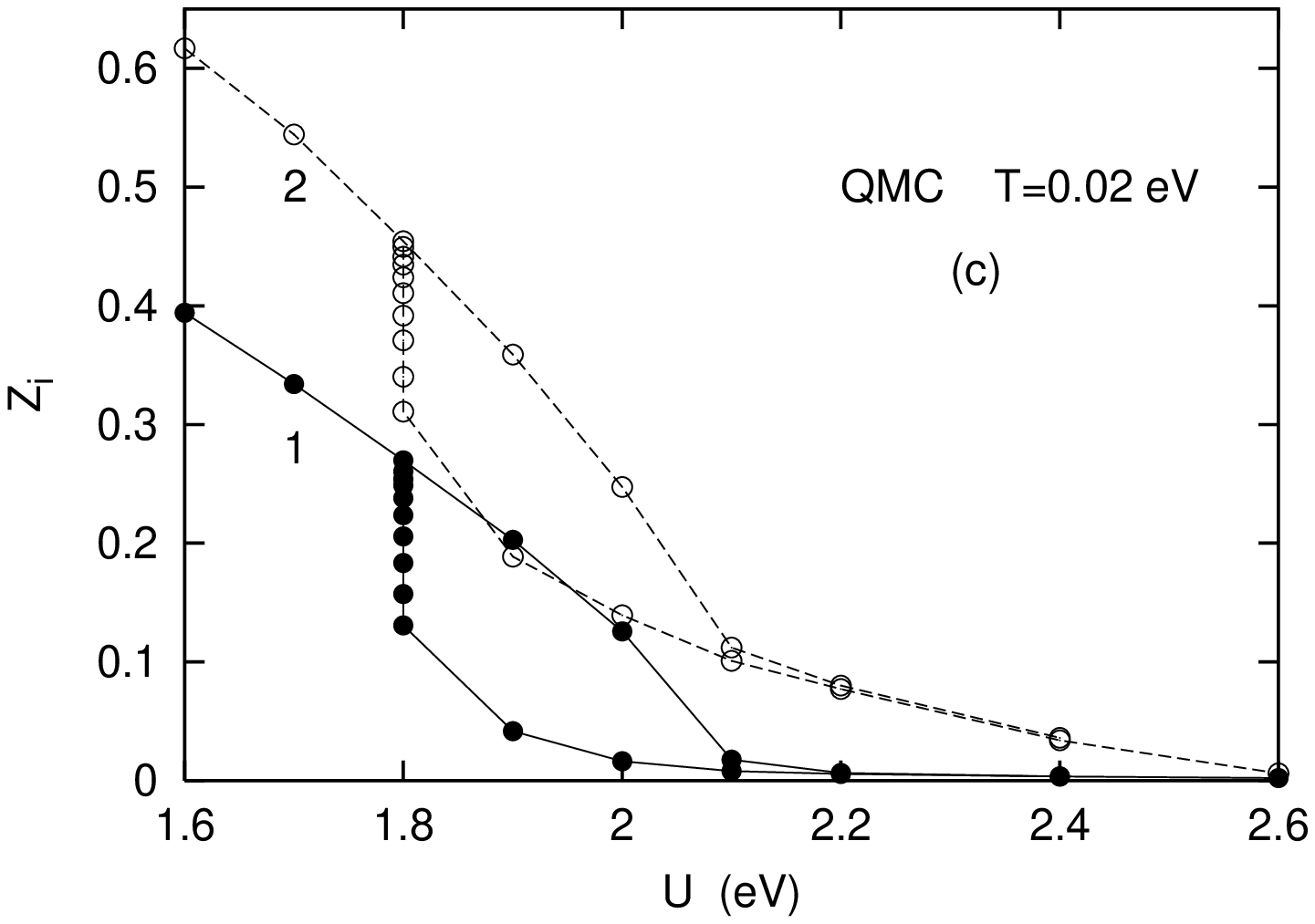}
  \end{center}
   \vskip-3mm
\caption{
Quasi-particle weights $Z_i$ of two-band Hubbard models as  
functions of on-site Coulomb energy $U$ calculated within the 
QMC-DMFT for $J=U/4$. (a) $T=0.125$~eV, (b) $T=0.031$~eV:
Solid dots: isotropic two-band models with degenerate narrow bands 
$W_i=2$~eV ($11$) or wide bands $W_i=4$~eV ($22$); open dots:  
non-isotropic two-band model with $W_2=2W_1=4$~eV ($12$). 
Symbols ($+$) in(b): results for $T=0.020$~eV.
(c) Hysteresis obtained for non-isotropic model at $T=0.020$~eV: 
solid (open) dots: narrow (wide) band; upper (lower) points: 
increasing (decreasing) $U$. Lines are guides to the eye.   
}\end{figure}

Fig.~9 shows the quasi-particle weights $Z_i$ as functions of $U$ 
obtained from analogous QMC calculations for $J=U/4$, $U'=U/2$. 
In the cross-over region at high temperatures ($T=0.125$~eV) the 
larger anisotropy caused by the stronger Hund's rule coupling leads 
to a pronounced overall broadening of the transition region. 
This applies also to the isotropic two-band models with degenerate 
narrow or wide subbands. Nevertheless, in the anisotropic case 
($12$) the $Z_i$ approach one another with increasing $U$ and 
exhibit qualitatively similar asymptotic behavior. 

The results for $T=0.031$~eV shown in Fig.~9(b) exhibit much 
sharper transitions in the isotropic as well as non-isotropic 
two-band systems. (In the single-band case\cite{DMFT,bulla} 
\,$T_c^{\rm QMC}/T_c^{\rm IPT}\approx0.75$. Assuming the same 
ratio for the two-band case we estimate from Fig.~5 for the 
non-isotropic model: $T_c^{\rm QMC}\approx 0.038$~eV.)
In the non-isotropic case the overall behavior of the 
quasi-particle weights is remarkably similar to the IPT results 
discussed in the previous section. While $Z_1$ drops almost to 
zero near $U_{c2}=2.1$~eV $Z_2$ drops to a finite value and 
then decreases gradually towards larger $U$. The characteristic 
kink in $Z_2$ becomes even more pronounced at $T=0.02$~eV (see 
$+$ symbols). This change in slope of $Z_2(U)$ near $U_{c2}$ 
suggests that both subbands undergo a common transition at the 
same Coulomb energy and that the wide subband exhibits 
pronounced `bad-metal' behavior above $U_c$. 

The hysteresis behavior of the $Z_i$ for $T=0.02$~eV is shown 
more detail in Fig.~9(c). While for increasing $U$ both $Z_i$ 
exhibit a change of slope near $U_{c2}\approx2.1$~eV, for
decreasing $U$ the common transition occurs at $U_{c1}\approx1.8$~eV.         
(The vertical line of dots at 1.8~eV indicates the iterations from  
the lower to the upper branch.) As in the case of the IPT, there is 
no evidence for a second transition in the wide subband at larger $U$.  

As can be seen in Fig.~10, the self-energies $\Sigma_i(i\omega_n)$ 
close to the critical Coulomb energy derived within the QMC method 
are consistent with those obtained in the IPT (see Fig.~6). At 
$U\leq2.4$~eV  the wide band is in the cross-over region between
metallic and insulating behavior. For $U=2.7$~eV $\Sigma_2(i\omega_n)$
becomes inversely proportional to $\omega_n$, i.e., a gap opens up.
The narrow band undergoes a similar cross-over behavior, except at 
slightly lower values of $U$.    
\begin{figure}[t!]
  \begin{center}
  \includegraphics[width=5.0cm,height=8cm,angle=-90]{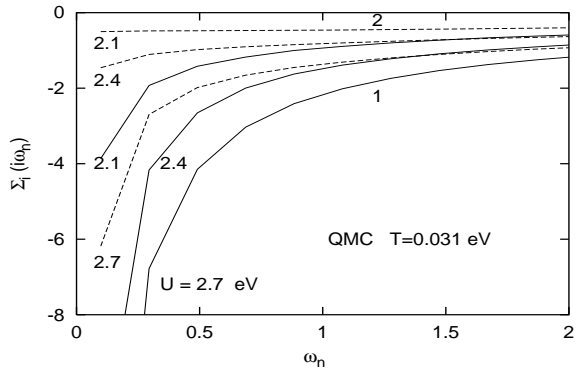}
  \end{center}
\caption{
Self-energy $\Sigma_i(i\omega_n)$ for two-band Hubbard model at 
$T=0.031$~eV near critical $U$ at small Matsubara frequencies,
calculated within QMC-DMFT. 
Solid (dashed) curves: narrow (wide) subband.
}\end{figure}

Fig.~11 shows the quasi-particle spectra for $T=0.031$~eV in the 
vicinity of the critical Coulomb energy. When the narrow band is 
about to open a gap, the wide band is still fairly metallic. Its
own gap would be even smaller and is therefore readily filled by
finite temperature tails extending from the Hubbard peaks towards
the Fermi level. Only if $U$ is increased to about 2.7~eV is the
gap in the wide band large enough to not be obliterated by this
temperature broadening.      

\begin{figure}[t!]
  \begin{center}
  \includegraphics[width=4.5cm,height=7cm,angle=-90]{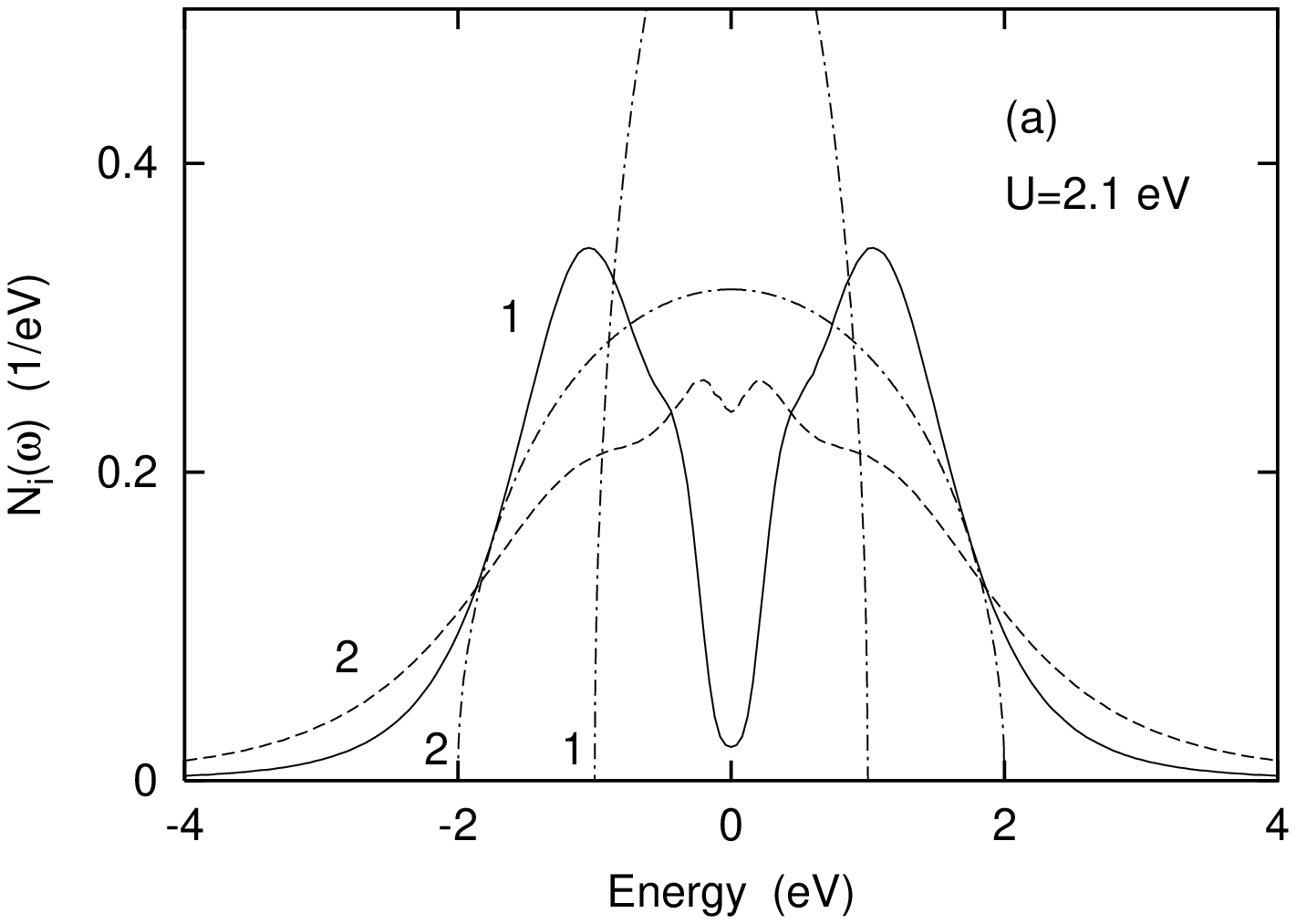}
  \includegraphics[width=4.5cm,height=7cm,angle=-90]{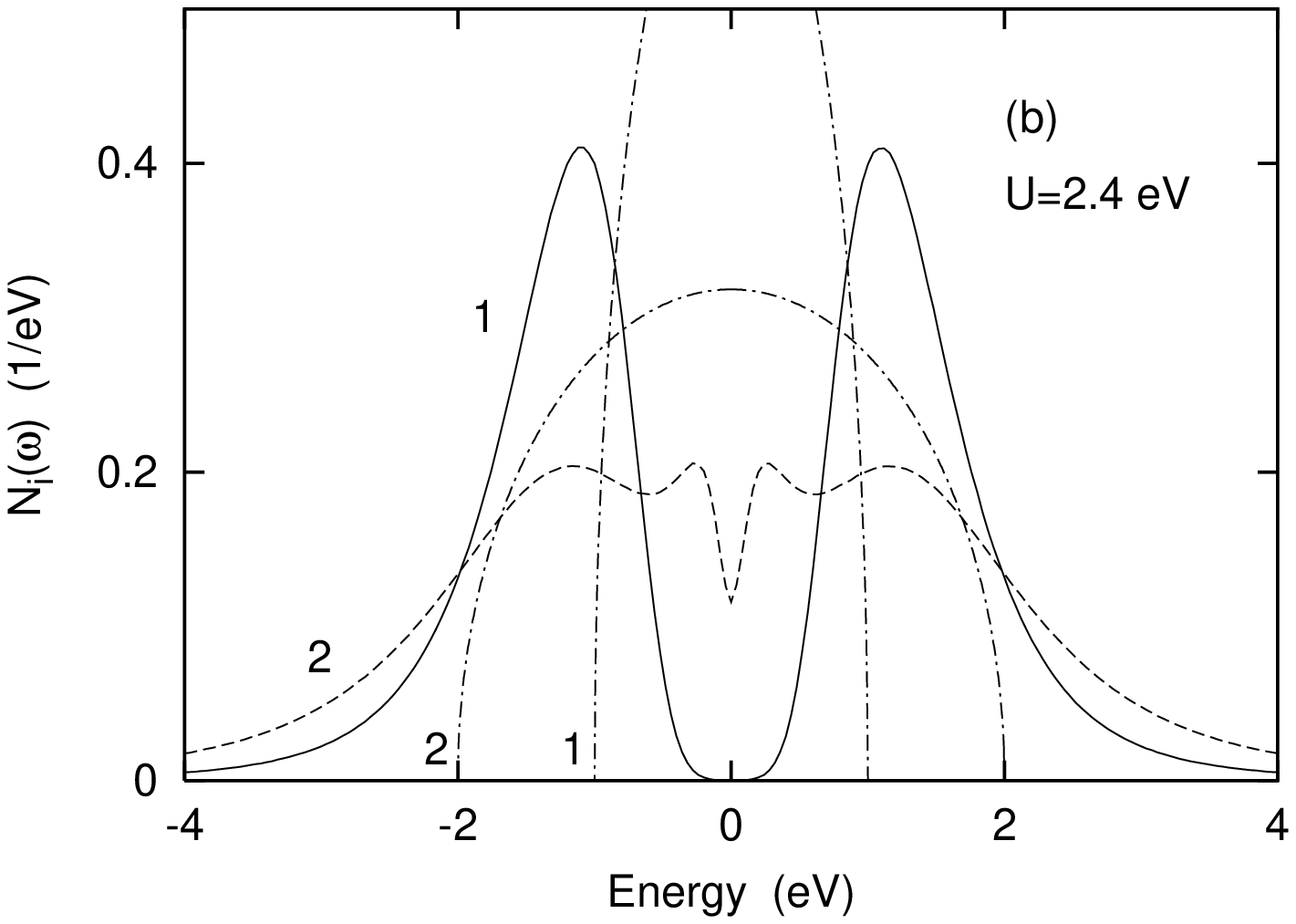}
  \includegraphics[width=4.5cm,height=7cm,angle=-90]{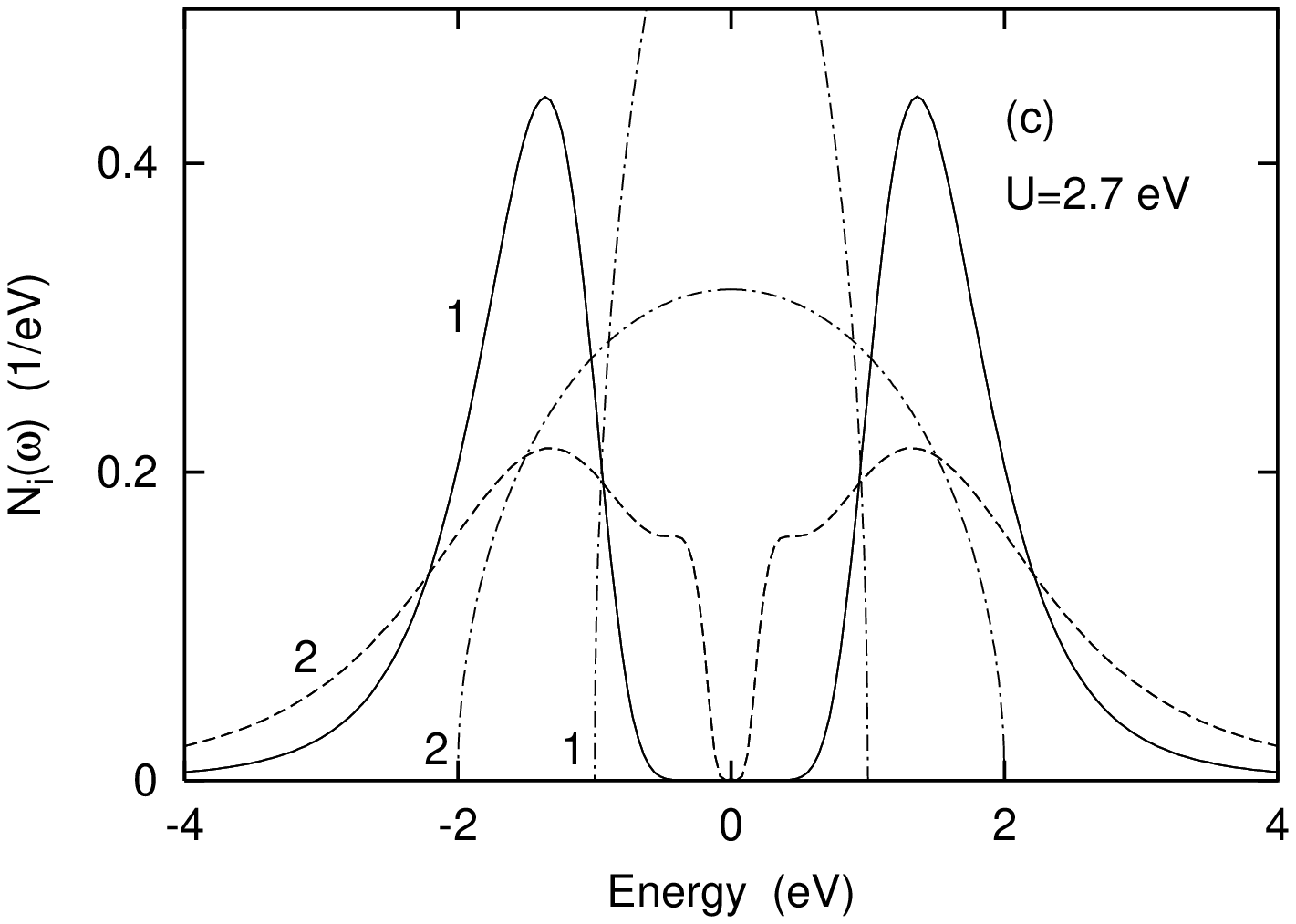}
  \end{center}
\caption{
Quasi-particle spectra $N_i(\omega)$ for two-band Hubbard model 
at $T=0.031$~eV, calculated within the QMC-DMFT. 
(a) $U=2.1$~eV; (b) $U=2.4$~eV; (c) $U=2.7$~eV.  
Solid (dashed) curves: narrow (wide) subband; 
dash-dotted curves: bare densities of states. 
}\end{figure}

The variation of $Z_i(U)$ with increasing $U$ shown in Fig.~9(b) is 
qualitatively similar to the one obtained by Koga et al.\cite{koga} 
within the exact diagonalization treatment of the non-isotropic 
two-band model: Near $U_c\approx2.6$~eV $Z_1$ practically vanishes 
while $Z_2(U)$ remains finite but shows a characteristic change in 
slope. At larger Coulomb energies $Z_2$ 
decreases gradually without any clear evidence for a subsequent metal 
insulator transition. ($U_c$ is somewhat larger than in our QMC
results since the ED calculations are carried out at near zero 
temperature. According to the phase diagram shown in Fig.~5
$U_{c2}(T)$ increases slightly with decreasing $T$.) 

Since the ED-DMFT results are basically similar to our QMC-DMFT 
calculations, the different conclusions reached in Refs.~11 and 
and 13 seem to stem primarily from different 
interpretations of the underlying physics: In our view
there is only one first-order transition, but with complex
quasi-metallic and -insulating subband features just below and
above this transition. This picture is based on the simultaneous 
discontinuous changes of the subband quasi-particle weights 
derived within the IPT-DMFT and the consistency between the IPT
and QMC results. On the other hand, if the vanishing of the $Z_i$ 
is used as a criterion, the subbands indeed behave very differently.
Whereas for the narrow band a critical Coulomb energy can readily 
be identified,  $Z_2$ falls off gradually and becomes negligibly
small at much higher values of $U$. This metallic `tail' of $Z_2$ 
makes it difficult to identify a second critical Coulomb energy. 
According to our QMC and IPT results, this metallic tail should
diminish at low temperature. It is not clear at present why the
ED results in Ref.~13 show such a tail although they 
apply to near zero temperature. Also, it is not clear to what
extent the different subband $U_c$'s obtained within the two-site 
DMFT depend on the minimal one-level representation of the
subband baths. The energy discretization inherent in both the ED
and two-site methods plays a crucial role close to the transition 
when dynamical correlation effects on small energy scales are 
particularly important. It would be interesting to perform ED 
and 2-site DMFT calculations in a wider range of temperatures 
in order to analyze the transition region in more detail.     
              
\section{Summary}

The nature of the Mott transition in multi-band systems 
was investigated within the dynamical mean field theory. Although in
practice subbands of transition metal oxides are usually coupled via 
single- as well as many-electron interactions, we focus here on the
effect of inter-orbital matrix elements of the Coulomb energy, thus
neglecting one-electron hybridization. In contrast to isotropic 
systems consisting of identical subbands, the metal insulator 
transition in materials with coexisting narrow and wide subbands
turns out to be remarkably complex. The present results confirm
our previous finding, namely, the existence of a single transition
rather than a sequence of orbital-selective transitions as the
on-site Coulomb energy is increased. Nevertheless, at low but finite 
temperatures below the transition the narrow band reveals a narrow 
coherent peak nearly separated from the Hubbard bands. Conversely, 
above the transition the wide band shows pronounced bad-metal 
behavior. This coexistence of quasi-metallic and -insulating 
spectral features in the vicinity of the transition makes the 
interpretation of the DMFT results non-trivial and partly explains 
the contradictory findings in Refs.~11 and 13.  

Since it is difficult to perform QMC calculations at very low 
temperatures, we have carried out extensive DMFT calculations within 
the IPT. The phase diagram is consistent with earlier results obtained 
for single-band models. In particular, the quasi-particle weights show 
the typical hysteresis behavior giving rise to two first-order lines 
$U_{c1}(T)$ and $U_{c2}(T)$ defining the stability boundaries of 
the insulating and metallic regions, respectively. 
For $T\leq0.01$~eV the results suggest that above the transition 
the metallic states in the wide band should disappear, i.e., that 
the excitation gaps in both bands should open at the same $U_c(T)$.    
At intermediate temperatures below $T_c$ the IPT results are 
consistent with the QMC calculations: The quasi-particle weights
in both subbands undergo simultaneous first-order transitions at 
$U_{c1}(T)$ or $U_{c2}(T)$. These transitions, however, are not
complete in one or the other of the subbands. For instance, the 
wide band shows appreciable metallicity above the transition. 
Similarly, the narrow band exhibits partly insulating Hubbard 
peaks below the transition. This mixture of metallic and 
insulating behavior diminishes towards low temperatures, i.e.,
purely metallic or insulating phases should evolve in both 
subbands.   
         
I like to thank A. Bringer for useful discussions.  
I also thank V. Oudovenko for sharing his subroutines 
for imaginary time/frequency Fourier transforms.    

Email: a.liebsch@fz-juelich.de


\begin{thebibliography}{99}
\bibitem{dip}  
   The tiny dips near $E_F$ are presumably not 
   physical and stem from uncertainties of the
   maximum entropy method.
\end{thebibliography}
 \end{document}